\documentclass[a4paper,11pt]{article}
\pdfoutput=1 

\usepackage{jheppub} 

\usepackage[T1]{fontenc} 
\usepackage{float}
\usepackage{epstopdf}
\usepackage{amsmath}
\makeatletter
\gdef\@fpheader{}
\makeatother

\title{\boldmath Magnetized hairy black holes of dimensionally continued gravity coupled to double-logarithmic electrodynamics}

\author{Askar Ali} 
\affiliation{Department of Mathematics, Quaid-i-Azam University, Islamabad, Pakistan}

\emailAdd{askarali@math.qau.edu.pk}

\abstract{A recently proposed model for nonlinear electrodynamics has been minimally coupled to dimensionally continued gravity and the topological black holes in the presence of conformal scalar field were studied. In this set up, the new magnetized hairy black hole solution has been found and its thermodynamic properties have also been analyzed. The exact expressions for mass, entropy, Hawking temperature and heat capacity are derived and local thermodynamic stability for the resulting black holes has been checked. In addition to this, the modified Smarr's formula is constructed and the generalized first law has also been verified. Finally, the hairy magnetized black holes in general Lovelock-scalar gravity have also been studied.
\vspace{80 mm}
}

\notoc 

\begin{document}
\maketitle
\flushbottom 


\section{Background}
\label{sec:intro}
The Einstein's general relativity (EGR) is a non-renormalizable theory, however, it is believed that a renormalizable theory \cite{1} can be constructed if one makes the higher derivative corrections in it. Therefore, it can be highly motivated to consider higher order gravity theories. Such higher order gravities are providing a useful way for the investigation of universe acceleration without using the inclusion of dark energy. Among these theories, one is the well known Lovelock theory of gravity (LTG) \cite{2}. This theory contains dimensionally continued Euler characteristics and the EGR can be recovered from it in four dimensions. It is the unique modified theory which retains the second order character of the equation of motion. Thus, it is a ghost-free theory at a linear level. It has been proven that among others, the second order Lovelock term is also relevant in the low energy limit of the string theory \cite{3,4}. This indicates that the LTG may forms the high energy generalization for the classical gravity. The arbitrariness for the choice of coupling coefficients in the Lovelock action produces difficulties in the determination of the explicit solution. Therefore, Ba\~{n}ados, Teitelboim and Zanelli introduced a suitable choice for these Lovelock coefficients which enables us to obtain the explicit solution. Based upon this choice, the new theory known as dimensionally continued gravity (DCG) \cite{5} has been developed. Vacuum and charged black hole solutions in this theory have been found in Refs. \cite{5,6,7}. Thermodynamic properties and stability analysis of these dimensionally continued black holes were also studied in Refs. \cite{5,6,7,7a,8,9,10,10b}.

 Since, the Lovelock action contains the nonlinear terms of curvature invariants, therefore, it looks more natural to take the nonlinear sources in the matter sector. When one wants the electromagnetic field as a source of gravitational field then it is much motivating to coupled nonlinear electrodynamics (NLED) with Lovelock gravity. The idea of NLED is important because the linear Maxwell electrodynamics cannot explain the electromagnetic phenomenon very well. In 1936, Heisenberg and Euler formulated a model of NLED for the description of quantum electrodynamics \cite{11}. Similarly, Born and Infeld also formulated a new nonlinear theory of electrodynamics in which the self-energy of electrons gives a finite value. This model is known as Born-Infeld (BI) electrodynamics \cite{12}. Later, it was also shown that this model can be reemerged in string theory as an effective action describing the abelian vector field coupled to a virtual open Boson string \cite{13,14}. BI formalism is also very useful in studying dark energy, holographic superconductors and holographic entanglement entropy \cite{15,16,17}, etc. Note that unlike Maxwell's theory, the electric field and potential corresponding to the charged particle are finite in this model. Therefore, it would be very motivating if one coupled BI electrodynamics with gravity theory and investigate the black holes. The first solution in this context was first derived by Hoffmann in Ref. \cite{18}. It has also been shown that this solution is free of curvature singularity at $r=0$. In addition to this, the different static black hole solutions with or without cosmological constant in the presence of BI electromagnetic field sources were derived in Refs. \cite{19,20,21,22,23,24,25}. Thermodynamic and physical properties of BI black holes were also explored in Refs. \cite{27,28,29,30}. There also exists several other models of NLED other than the BI model, for example, the models introduced in Refs. \cite{31,32,33,34,35}. In the framework of these models, different static black holes and their physical properties were studied in Refs. \cite{36,37,38,39,40,41,42,43,44,45,46,47,48,49,50,51,52,53,54,55,56,57,58,59,60,61,62,63,64,65,66,67,68,69,72}. Recently, the thermodynamic and physical properties corresponding to NLED black holes of DCG were also studied \cite{73,74}. In addition to spherically symmetric solutions, the rotating charged Gauss-Bonnet black brane solutions have also been obtained in Ref. \cite{74a}. Furthermore, dilatonic black holes with NLED were investigated in Refs. \cite{74b,74c}. Recently, charged black holes surrounded by dark fluid \cite{74d} and with quintessence in a string-cloud background \cite{74e} have also been studied.
 
 Investigations related to no-hair theorems have both similarly long history and active present. These famous theorems disallow the existence of asymptotically flat black holes of EGR in the presence of conformal scalar field \cite{85,86}. When cosmological constant is zero, the conformal hairy black holes in four spacetime dimensions can be found but the scalar field appears to be infinite at the event horizon \cite{87}. It has been shown that the higher dimensional generalization to these black holes do not exist \cite{88}. However, in the presence of cosmological constant, both the three and four dimensional conformal hairy black holes with the scalar field analytic on the event horizon were determined in Refs. \cite{89,90}. However, until recently higher dimensional black holes in this regard were not determine when no-go results were reported \cite{91}. Recently, the new gravity coupled to conformal scalar field has been developed in Refs. \cite{92,93}. The action function for this theory can be written as
 \begin{equation}
 I_S=\int d^dx\sqrt{-g}\sum_{p=0}^{n-1}\bigg(b_p\phi^{d-4p}\delta^{\mu_1...\mu_{2p}}_{\nu_1...\nu_{2p}}S^{\nu_1\nu_2}_{\mu_1\mu_2}...S^{\nu_{2p-1}\nu{2p}}_{\mu_{2p-1}\mu{2p}}\bigg),\label{1}
 \end{equation}
 where $\delta^{\mu_1...\mu_{2p}}_{\nu_1...\nu_{2p}}$ denotes the generalized Kronecker delta while
 
 \begin{equation}\begin{split}
 S^{\gamma\alpha}_{\mu\nu}&=\phi^2R^{\gamma\alpha}_{\mu\nu}-2\delta^{[\gamma}_{[\mu} \delta^{\alpha]}_{\nu]}\nabla_{\rho}\phi\nabla^{\rho}\phi-4\phi\delta^{[\gamma}_{[\mu}\nabla_{\nu]}\nabla^{\alpha]}\phi\\&+8\delta^{[\gamma}_{[\mu}\nabla_{\nu]}\phi\nabla^{\alpha]}\phi.\label{2}\end{split}
 \end{equation}
  
  This is the most general form of the action describing the ghost-free theory. It is important to note that the LTG coupled to the conformal scalar field is also free of ghosts as the equations of motion for the metric and scalar field are still second order. In four dimensions or when $p=1$, the action (\ref{1}) describes the conformal scalar field gravity with the potential $\mathcal{V}(\phi)=\frac{\lambda}{4!}\phi^4$ and non-minimal coupling term $(-1/12)R\phi^2$ \cite{93}. Solutions representing higher dimensional hairy black holes were derived in Refs. \cite{93,94,95,96,97}. These black holes are considered as the first ones for $d>4$ spacetime dimensions. It has been shown that the hairy black holes are more thermodynamically stable and physical \cite{94,95}. The stability analysis of the hairy charged black holes in Einstein-Maxwell theory has also been performed by employing the radial perturbations \cite{95a,95b}. By using the similar approach, the stability analysis of Einstein-Gauss-Bonnet black holes has been investigated in Ref. \cite{95c}. In addition to this, the linear stability of black holes in quadratic gravity were discussed in Ref. \cite{95d}. The thermodynamic stability and thermal phase transitions of the hairy black holes in Born-Infeld-Lovelock gravity were studied in Ref. \cite{98}. Similarly, the magnetized hairy black holes of DCG and their thermodynamics were also studied in Ref. \cite{74}.  Therefore, in this work I am also studying the magnetized hairy black holes of DCG coupled to double-logarithmic electrodynamics. Black holes are considered as one of the most fascinating objects which exhibit many interesting physical consequences. Bekenstein and Hawking showed that the entropy of black hole is proportional to the horizon area and its temperature is given by the surface gravity evaluated on the event horizon \cite{99,100,101,102,103}. In 1980's, Hawking and Page showed that there can be possible phase transitions between AdS Schwarzschild black hole and thermal AdS space \cite{104}. The AdS/CFT correspondence says that Hawking-Page transition appeared as the gravitational dual for the confinement/deconfinement phase transition \cite{105,106,107,108}. Thus, due to the importance of thermodynamical aspects in black holes, I have also calculated the exact expressions for thermodynamic quantities associated to the obtained hairy black holes in this paper. 
  
  The plan of the paper is as follows. In Section 2, I give the basic formulations of dimensionally continued gravity and double-logarithmic electrodynamics. I obtain the field equations in this setup and derive the new magnetized hairy solution. In Section 3, I present the different thermodynamic quantities in terms of the event horizon and magnetic monopole charge. The generalized first law and Smarr's formula have also been constructed. Section 4 is devoted to the investigation of magnetized hairy black holes in general Lovelock gravity. Finally, I present a brief conclusion of the paper in section 5.

\section{Magnetized hairy dimensionally continued black holes} 

The action describing the Lovelock gravity coupled to conformal scalar field and double-Logarithmic NLED is given by
\begin{equation} \begin{split}
I&=\frac{1}{16\pi G}\int d^dx\sqrt{-g}\bigg[\sum_{p=0}^{n-1}\frac{1}{2^p}\delta^{\mu_1...\mu_{2p}}_{\nu_1...\nu_{2p}}\bigg(a_p R^{\nu_1\nu_2}_{\mu_1\mu_2}...R^{\nu_{2p-1}\nu_{2p}}_{\mu_{2p-1}\mu_{2p}}+16\pi G b_p\phi^{d-4p}\\&\times S^{\nu_1\nu_2}_{\mu_1\mu_2}...S^{\nu_{2p-1}\nu_{2p}}_{\mu_{2p-1}\mu_{2p}}\bigg)+4\pi GL_M(P)\bigg],
\label{3}\end{split}
\end{equation}
where coefficients $a_p$ and $b_p$ are arbitrary constants. Also $G$ is the Newtonian constant, $R^{\alpha\beta}_{\mu\nu}$ are the curvature tensor components and $S^{\alpha\beta}_{\mu\nu}$ are the components of the 4th rank tensor given by (\ref{2}). Note that, the tensor $S^{\alpha\beta}_{\mu\nu}$ transforms homogeneously under the conformal transformation defined by $g_{\mu\nu}\rightarrow \Omega^2g_{\mu\nu}$, $\phi\rightarrow\Omega^{-1}\phi$ when $S^{\alpha\beta}_{\mu\nu}\rightarrow\Omega^{-4}S^{\alpha\beta}_{\mu\nu}$. Furthermore, $L_M(P)$ refers to the Lagrangian density of double-Logarithmic electrodynamics and is given as 
\begin{equation}\begin{split}
L_M(P)&=\frac{1}{2\beta}\bigg[\bigg(1-\sqrt{-2\beta P}\bigg)\log{\bigg(1-\sqrt{-2\beta P}\bigg)}+\bigg(1+\sqrt{-2\beta P}\bigg)\\&\times\log{\bigg(1+\sqrt{-2\beta P}\bigg)}\bigg],    \label{4}\end{split}
\end{equation}
where $P=F_{\mu\nu}F^{\mu\nu}=2\big(\textbf{B}^2-\textbf{E}^2\big)$ in which $\textbf{E}$ represents the electric field, $\textbf{B}$ is the magnetic field and $F_{\mu\nu}$ denotes the Maxwell tensor which in terms of gauge potential $A_{\mu}$ can be defined as $F_{\mu\nu}=\partial_{\mu}A_{\nu}-\partial_{\nu}A_{\mu}$. The action function (\ref{3}) describes dimensionally continued gravity when the coefficients $a_p$ are define as \cite{5,7a}
\begin{eqnarray}
a_p=\left( \begin{array}{c} n-1 \\ p \end{array} \right)\frac{(d-2p-1)!}{(d-2)!l^{2(n-p-1)}}. \label{5}
\end{eqnarray} 
The equations of motion corresponding to LTG can be obtained from the variation of action (\ref{3}) with respect to the metric tensor $g_{\mu\nu}$ as
\begin{equation}
-\sum_{p=0}^{s}\frac{a_p}{2^{p+1}}\delta^{\nu\lambda_1...\lambda_{2p}}_{\mu\rho_1...\rho_{2p}} R^{\rho_1\rho_2}_{\lambda_1\lambda_2}...R^{\rho_{2p-1}\rho_{2p}}_{\lambda_{2p-1}\lambda_{2p}}=16\pi GT^{(M)\nu}_{\mu}+16\pi GT^{(S)\nu}_{\mu},
\label{6}
\end{equation}
where the components of matter tensor $T^{(M)\nu}_{\mu}$ associated to the double-logarithmic electromagnetic field are given by
\begin{equation}\begin{split}
T^{(M)}_{\mu\nu}&=\frac{1}{2\beta}\bigg[\bigg(1-\sqrt{-2\beta P}\bigg)\log{\bigg(1-\sqrt{-2\beta P}\bigg)}+\bigg(1+\sqrt{-2\beta P}\bigg)\\&\times\log{\bigg(1+\sqrt{-2\beta P}\bigg)}\bigg]g_{\mu\nu}-\frac{2F_{\mu\lambda}F^{\lambda}_{\nu}}{\sqrt{-2\beta P}}\log{\bigg(\frac{1-\sqrt{-2\beta P}}{1+\sqrt{-2\beta P}}\bigg)}.\label{7}\end{split}
\end{equation}
Similarly, the components $T^{(S)\nu}_{\mu}$ correspond to the matter tensor of the conformal scalar field and can be obtained as
\begin{equation}\begin{split}
T^{(S)\nu}_{\mu}&=\sum_{p=0}^{n-1}\frac{b_p}{2^{p+1}}\phi^{d-4p}\delta^{\nu\lambda_1...\lambda_{2p}}_{\mu\rho_1...\rho_{2p}}S^{\rho_1\rho_2}_{\lambda_1\lambda_2}...S^{\rho_{2p-1}\rho_{2p}}_{\lambda_{2p-1}\lambda_{2p}}.\label{8}\end{split}
\end{equation}
The equations of motion for the double-logarithmic electromagnetic field can be obtained from the variation of Eq. (\ref{3}) with respect to gauge potential $A_{\mu}$ as
  \begin{equation}\begin{split}
 \partial^{\mu}\bigg[\frac{\sqrt{-g}}{\sqrt{-2\beta P}}\log{\bigg(\frac{1-\sqrt{-2\beta P}}{1+\sqrt{-2\beta P}}\bigg)}F_{\mu\nu}\bigg]=0.  \label{9}\end{split}
 \end{equation}
 It should be noted that by imposing the limit $\beta\rightarrow 0$ in the Lagrangian density (\ref{4}), one can attains the formulations of Maxwell's theory. Similarly, variation of the Eq. (\ref{3}) with respect to scalar field yields the equations of motion
 \begin{equation}\begin{split}
 \sum_{p=0}^{n-1}\frac{(d-2p)b_p}{2^p}\phi^{d-4p-1}\delta^{\lambda_1...\lambda_{2p}}_{\rho_1...\rho_{2p}}S^{\rho_1\rho_2}_{\lambda_1\lambda_2}...S^{\rho_{2p-1}\rho_{2p}}_{\lambda_{2p-1}\lambda_{2p}}=0.\label{10}\end{split}
 \end{equation}
 It can be noted that the above equation (\ref{10}) makes the trace of $T^{(S)\nu}_{\mu}$ equal to zero. Hence, this confirms the conformal coupling of scalar field with gravity. 
 In order to derive the static spherically symmetric solution, it is convenient to consider the general metric ansatz as
 \begin{equation}
 ds^2=-f(r)dt^2+\frac{dr^2}{f(r)}+r^2d\Sigma^2_{(\alpha)d-2},
 \label{11}
 \end{equation}
 where $d\Sigma^2_{(\alpha)d-2}$ defines the line element of $(d-2)$-dimensional hyper-surface of constant curvature equal to $(d^2-5d+6)\alpha$ such that $\alpha$ is a constant and takes the values $\alpha=0,+1,-1$, associated to flat, spherical and hyperbolic horizon topologies, respectively. The volume of this submanifold is represented by $\Sigma^{(\alpha)}_{d-2}$, which for the case of spherical horizon topology becomes $\Sigma^{+1}_{d-2}=\frac{2\pi^{(d-1)/2}}{\Gamma[(d-1)/2]}$.
 Now, for obtaining a solution which describes the pure magnetized hairy black hole, the choice $\textbf{B}\neq0$ and $\textbf{E}=0$ should be taken into account. Due to this assumption, the Maxwell's invariant can be expressed as $P=\frac{2Q^2}{r^{2d-4}}$ in which $Q$ represents the magnetic monopole charge. Furthermore, when the scalar field configuration is selected as
 \begin{eqnarray}
 \phi(r)=\frac{N}{r},\label{12}
 \end{eqnarray}
then the real valued function $\phi(r)$ satisfies the scalar field equations (\ref{10}) if the following equations hold for the parameter $N$
\begin{eqnarray}
\sum_{p=1}^{n-1}pb_p\frac{(d-1)!}{(d-2p-1)!}N^{2-2p}=0,\label{13}
\end{eqnarray}
 and 
 \begin{eqnarray}
 \sum_{p=1}^{n-1}b_p\frac{(d-1)!(d^2-d+4p^2)}{(d-2p-1)!}N^{-2p}=0.\label{14}
 \end{eqnarray}
 Since, there are two equations for the one unknown $N$. So, this implies that one of the equation is playing the role of constraint on the conformal coupling constants $b_p$'s. Thus, by considering the pure double-logarithmic magnetic field along with Eqs. (\ref{11})-(\ref{14}) for the conformal scalar field configuration, one can get the equation of motion (\ref{6}) as
 \begin{eqnarray}\begin{split}
 \frac{d}{dr}\bigg[\sum_{p=0}^{n-1}&\frac{(d-2)!a_pr^{d-1}}{2(d-2p-1)!}\bigg(\frac{\alpha-f(r)}{r^2}\bigg)^p\bigg]=-\frac{16\pi Gr^{d-2}}{\beta}\log{\bigg(1+\frac{4\beta Q^2}{r^{2d-4}}\bigg)}\\&+64\pi G Q\sqrt{\beta}\arctan{\bigg(\frac{2Q\sqrt{\beta}}{r^{d-2}}\bigg)}+16\pi G \sum_{p=0}^{n-1}\frac{b_p(d-2)!N^{d-2p}}{(d-2p-2)!r^2}.\label{14a}\end{split}
 \end{eqnarray}
 Now, by taking the values of coupling coefficients $a_p$ given in (\ref{5}), the above equation can be written in the more simplified form as
 \begin{eqnarray}\begin{split}
 \frac{d}{dr}\bigg[r^{d-1}\bigg(\frac{1}{l^2}&+\frac{\alpha-f(r)}{r^2}\bigg)^{n-1}\bigg]=-\frac{32\pi Gr^{d-2}}{\beta}\log{\bigg(1+\frac{4\beta Q^2}{r^{2d-4}}\bigg)}\\&+128\pi G Q\sqrt{\beta}\arctan{\bigg(\frac{2Q\sqrt{\beta}}{r^{d-2}}\bigg)}+32\pi G \sum_{p=0}^{n-1}\frac{b_p(d-2)!N^{d-2p}}{(d-2p-2)!r^2}.\label{14b}\end{split}
 \end{eqnarray}
 Integration of this equation gives the solution as
 \begin{eqnarray}\begin{split}
 f(r)&=\alpha+\frac{r^2}{l^2}-r^2\bigg(\frac{16\pi G M}{\Sigma_{d-2}^{(\alpha)}r^{d-1}}+\frac{\delta_{d,2n-1}}{r^{d-1}}+\frac{32\pi G H}{r^d}-\frac{16\pi G}{\beta (d-2)(d-1)^2}\\&\times\log{\bigg(1+\frac{4\beta Q^2}{r^{2d-4}}\bigg)}+\frac{64\pi G Q^2r^{2-d}}{\sqrt{\beta}(d-1)(d-2)(d-3)}\arctan{\bigg(\frac{2Q\sqrt{\beta}}{r^{d-2}}\bigg)}\\&-\frac{128\pi d Q^2G}{(d-3)(d-1)^2r^{2d-4}}F_1\bigg[1,\frac{d-3}{2d-4},\frac{3d-7}{2d-4},-\frac{4\beta Q^2}{r^{2d-4}}\bigg]\bigg)^{\frac{1}{n-1}},\label{15}\end{split}
 \end{eqnarray}
 where 
 \begin{eqnarray}
 H=\sum_{p=0}^{n-1}b_p\frac{(d-2)!N^{d-2p}}{(d-2p-2)!}.\label{16}
 \end{eqnarray}
 \begin{figure}[h]
 	\centering
 	\includegraphics[width=0.8\textwidth]{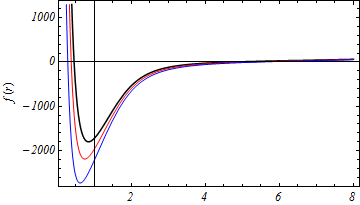}
 	\caption{$f(r)$ versus $r$. The parameters are selected as $Q=10$, $\delta_{d,2n-1}=5$, $\beta=0.1$, $\Sigma_{d-2}=100$, $l=1$, $H=1$, $n=2$, $d=5$, $M=10$ (black), $M=500$ (red) and $M=1000$ (blue).}\label{skr}
 \end{figure} 
 The constant of integration $M$ in Eq.(\ref{15}) is related to the mass of the black hole and the additive constant $\delta_{d,2n-1}$ is chosen such that in the limit $M\rightarrow0$, the black hole shrinks into a single point. Fig. \ref{skr} shows the behaviour of the metric function (\ref{15}) for a suitably chosen values of the parameters $Q$, $\beta$ and $H$. The value of radial coordinate $r$ at which the curve intersects the horizontal axis indicates the position of black hole's horizon. It is easy to see that by choosing $b_p=0$ or in the limit $H\rightarrow 0$, the new class of dimensionally continued topological black holes without scalar hair in the framework of double-Logarithmic electrodynamics can be obtained. The black hole solutions of DCG with Maxwell source can be regained in the weak field limit i.e. when $\beta\rightarrow 0$. Moreover, the neutral dimensionally continued black holes can be recovered when $Q=0$ and $H\rightarrow 0$. It is worthwhile to mentioned here that the hairy black hole solutions given by (\ref{15}) are possible only for $d\geq5$. Because the equations Eqs. (\ref{13})-(\ref{14}) hold in four dimensions only when all the coupling constants $b_p$'s are zeros. Hence, we cannot find the hairy black hole solutions in the spacetime dimensions $d=4$.
 
 The curvature invariants associated to the static ansatz (\ref{11}) can be define as
 
 \begin{eqnarray}\begin{split}
 R(r)&=\bigg[(d-2)(d-3)\bigg(\frac{\alpha-f(r)}{r^2}\bigg)-\frac{d^2f}{dr^2}-\frac{2(d-2)}{r}\frac{df}{dr}\bigg],\label{17}\end{split}
 \end{eqnarray}
 and
 \begin{eqnarray}\begin{split}
 K(r)&=\bigg[2(d-2)(d-3)\bigg(\frac{\alpha-f(r)}{r^2}\bigg)^2-\bigg(\frac{d^2f}{dr^2}\bigg)^2+\frac{2(d-2)}{r^2}\bigg(\frac{df}{dr}\bigg)^2\bigg].\label{18}\end{split}
 \end{eqnarray}
 So, by substituting the obtained solution (\ref{15}) into above curvature scalars, one can show that both of these invariants are infinite at the origin $r=0$. This confirms the existence of a true curvature singularity at this point. Hence, the resulting solution represents the magnetized hairy black hole.
 
 \section{Thermodynamics of magnetized hairy dimensionally continued black holes}
   
   In order to study the thermodynamic properties for black holes, one should need to calculate the basic thermodynamic quantities, for instance, Hawking temperature, entropy and heat capacity. The horizons of the black hole can be determined from the roots of $f(r)=0$. Thus, using this fact the finite mass of the black hole as a function of event horizon $r_+$ can be found as 
    \begin{eqnarray}\begin{split}
    M&=\frac{\Sigma_{d-2}}{16\pi G}\bigg[r_+^{d-1}\bigg(\frac{1}{r_+^2}+\frac{1}{l^2}\bigg)^{n-1}-\delta_{d,2n-1}-\frac{16\pi G H(d-2)}{r_+}\\&+\frac{128\pi GQ^2d}{(d-3)(d-1)^2r_+^{d-3}}F_1\bigg[1,\frac{d-3}{2d-4},\frac{3d-7}{2d-4},-\frac{4\beta Q^2}{r_+^{2d-4}}\bigg]+\frac{16\pi G r_+^{d-1}}{\beta(d-2)(d-1)^2}\\&\times\log{\bigg(1+\frac{4\beta Q^2}{r_+^{2d-4}}\bigg)}-\frac{64\pi GQ^2r_+}{\sqrt{\beta}(d-1)(d-2)(d-3)}\arctan{\bigg(\frac{2Q\sqrt{\beta}}{r_+^{d-2}}\bigg)}\bigg]. \label{19}\end{split}
    \end{eqnarray}
     \begin{figure}[h]
    	\centering
    	\includegraphics[width=0.8\textwidth]{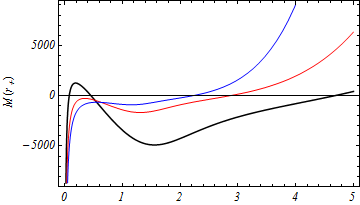}
    	\caption{Mass function $M$ versus $r_+$. The parameters are selected as $Q=10$, $\beta=0.1$, $\Sigma_{d-2}=100$, $l=1$, $H=1$, $n=2$, $d=5$ (black), $n=2$, $d=6$ (red) and $n=3$, $d=7$ (blue).}\label{skr1}
    \end{figure} 
    
    The behaviour of the mass as a function of horizon radius is shown in Fig. \ref{skr1}. It is shown that for the resulting hairy black holes given by (\ref{15}), there exist one or more horizons. Those values of $r_+$ for which the mass is positive indicate the existence of black holes with such horizon radii. However, those values of $r_+$ at which it is negative do not correspond to the horizons of black holes. 
    
    The Hawking temperature \cite{103} can be expressed through the definition of surface gravity $\kappa_s$ as $T_H=\kappa_s/2\pi$. Thus, associated to (\ref{15}), the temperature can be calculated as
     \begin{eqnarray}\begin{split}
   T_{H}&=\frac{r_+}{4\pi}\bigg[-\frac{2}{r_+^2}+\frac{r_+^{2n-3}l^{2n-4}}{(n-1)(r_+^2+l^2)^{n-2}}\bigg(\frac{(d-1)}{r_+}\bigg(\frac{1}{r_+^2}+\frac{1}{l^2}\bigg)^{n-1}+\frac{16\pi GH(d-2)}{r_+^{d+1}}\\&+\frac{16\pi G}{\beta(d-1)(d-2)r_+}\log{\bigg(1+\frac{4\beta Q^2}{r_+^{2d-4}}\bigg)}-\frac{128\pi G Q^2(d+1)}{(d-1)^2r_+(r_+^{2d-4}+4\beta Q^2)}\\&-\frac{64\pi GQ^2\arctan{\bigg(\frac{2Q\sqrt{\beta}}{r_+^{d-2}}\bigg)}}{\sqrt{\beta}(d-1)(d-2)(d-3)r_+^{d-1}}+\frac{128\pi G Q^3}{(d-1)(d-3)r_+(r_+^{2d-4}+4\beta Q^2)}\bigg)\bigg].\label{20}\end{split} 
    \end{eqnarray}
    \begin{figure}[h]
    	\centering
    	\includegraphics[width=0.8\textwidth]{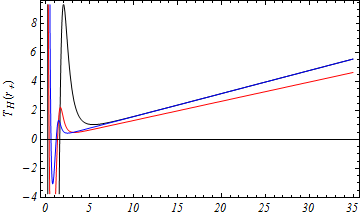}
    	\caption{Hawking temperature versus $r_+$. The parameters are selected as $Q=10$, $\beta=0.1$, $\Sigma_{d-2}=100$, $l=1$, $H=1$, $n=2$, $d=5$ (black), $n=2$, $d=6$ (red) and $n=3$, $d=7$ (blue).}\label{skr2}
    \end{figure} 

The graph of Hawking temperature is plotted in Fig. \ref{20} for different values of dimensionality parameter $d$. Those values of $r_+$ at which the temperature is negative describe the region of thermodynamic instability. The point at which the temperature changes sign corresponds to phase transition of type I and the region in which it is positive implies the black hole is physical.
   
   The entropy of the dimensionally continued black hole can be computed through using Wald's method \cite{109,110}. Thus, by employing this method, it can be expressed as
    \begin{equation}
    S=-2\pi\oint d^{d-2}x\sqrt{\gamma}\frac{\partial L}{\partial R_{abcd}}\epsilon_{ab}\epsilon_{cd}, \label{21}
    \end{equation}
    in which $L$ denotes the Lagrangian density of gravitational field, $\gamma$ is the determinant of the induced metric defined on the horizon and $\epsilon_{ab}$ represents the binormal to the horizon. Hence, the entropy for the dimensionally continued hairy black hole solution (\ref{15}) can be obtained as
    \begin{equation}\begin{split}
    S&=\frac{\Sigma_{d-2}}{4G}\sum_{p=1}^{n-1}\frac{pb_p(d-2!)N^{d-2p}}{(d-2p)!}+\frac{(n-1)\Sigma_{d-2}r_+^d}{4G(d-2n+2)}\bigg(\frac{1}{r_+^2}+\frac{1}{l^2}\bigg)^{n-1}\\&\times F_1\bigg[1,\frac{d}{2},\frac{d-2n+4}{2},\frac{-r_+^2}{l^2}\bigg]. \label{22}\end{split}
    \end{equation}
    The extended first law in this scenerio takes the form
      \begin{equation}\begin{split}
      dM&=T_HdS+\Phi_QdQ+W_{\beta}d\beta+\sum_{p=0}^{n-1}B^{(p)}db_p, \label{23}\end{split}
      \end{equation}
      where the conjugate quantity
      \begin{equation}\begin{split}
      \Phi_Q&=\frac{8\pi r_+\Sigma_{d-2}}{(d-3)(d-2)(d-1)^2}\bigg[\frac{(d^2-2d-3+Qd-Q)r_+^{d-2}}{(r_+^{2d-4}+4\beta Q^2)}-\frac{(d-1)}{\sqrt{\beta}}\arctan{\bigg(\frac{2Q\sqrt{\beta}}{r_+^{d-2}}\bigg)}\\&+\frac{d(d-1)}{r_+^{d-2}}F_1\bigg(1,\frac{d-3}{2d-4},\frac{3d-7}{2d-4},-\frac{4\beta Q^2}{r_+^{2d-4}}\bigg)\bigg], \label{24}\end{split}
      \end{equation}
      is the magnetic potential associated to magnetic charge $Q$. Similarly, the conjugate quantity relative to the parameter $\beta$ can be computed as
       \begin{equation}\begin{split}
      W_{\beta}&=\frac{\Sigma_{d-2}}{(d-2)(d-1)^2r_+\beta^2}\bigg[\frac{2Q^2\sqrt{\beta}r_+^2}{(d-3)}\bigg((d-1)\arctan{\bigg(\frac{2Q\sqrt{\beta}}{r_+^{d-2}}\bigg)}\\&+\frac{2\sqrt{\beta}r_+^{d-2}(d^2-2d-3+dQ+Q)}{r_+^{2d-4}+4\beta Q^2}\bigg)-r_+^d\log{\bigg(1+\frac{4\beta Q^2}{r_+^{2d-4}}\bigg)}\\&-\frac{4dQ^2\beta}{r_+^{2d-4}}F_1\bigg(1,\frac{3-d}{4-2d},\frac{7-3d}{4-2d},-\frac{4\beta Q^2}{r_+^{2d-4}}\bigg)\bigg]. \label{25}\end{split}
      \end{equation}
      While, the conjugate quantities $B^{(p)}$ associated to the coupling constants $b_p$ can be calculated as
       \begin{equation}\begin{split}
       B^{(p)}=-\frac{4\Sigma_{d-2}}{r_+}\sum_{p=0}^{n-1}\frac{(d-2)!N^{d-2p}}{(d-2p-2)!}. \label{26}\end{split}
       \end{equation}
       
       The generalized Smarr's relation associated to the above thermodynamic and conjugate quantities can be followed as
      \begin{equation}\begin{split}
      (d-3)M&=(d-2)T_HS+(d-3)\Phi_QQ+2\beta W_{\beta}d\beta+(d-2)\sum_{p=0}^{n-1}B^{(p)}b_p. \label{27}\end{split}
      \end{equation} 
      Since, it is clear from Eqs. (\ref{13}) and (\ref{14}) that $b_p's$ are not all independent. Thus, it should be emphasized that the variations of coupling constant $b_p$ in the first law (\ref{23}) are not all independent.
    
    The heat capacity is defined by the relation
    \begin{equation}
    C_Q=T_{H}(r_h)\frac{dS}{dT_H}|_{Q}. \label{28}
    \end{equation}
    
    Differentiation of Eq. (\ref{17}) with respect to $r_+$ gives
     \begin{equation}\begin{split}
    \frac{\partial T_H}{\partial r_+}&=\frac{1}{4\pi}\bigg[\Xi_1(r_+)+\Xi_2(r_+)+\Xi_3(r_+)+\Xi_4(r_+)+\frac{2}{l^2}\bigg], \label{29} \end{split}
    \end{equation}
    \begin{equation}\begin{split}
    \Xi_1(r_+)&=\frac{2(d-1)(2-n)r_+^2}{(n-1)}-2(d-1)l^2+\frac{(2n-3)(d-1)}{(n-1)(r_+^2+l^2)}\\&+\frac{16\pi GH(d-2)l^{2n-4}(r_+^2(1-d)+(2n-3-d)l^2)}{(n-1)r_+^{d-2n+4}(r_+^2+l^2)^{n-1}}, \label{30}\end{split}
    \end{equation}
    \begin{equation}\begin{split}
    \Xi_2(r_+)&=-\frac{16\pi G l^{2n-4}r_+^{2n-4}(r_+^2+l^2)^{1-n}}{\beta(d-2)(d-1)(n-1)(r_+^{2d-4}+4\beta Q^2)}\bigg[(8d-16)\beta Q^2(r_+^2+l^2)\\&-(r_+^2+(2n-3)l^2)(r_+^{2d-4}+4\beta Q^2)\log{\bigg(1+\frac{4\beta Q^2}{r_+^{2d-4}}\bigg)}\bigg], \label{31} \end{split}
    \end{equation}
    \begin{equation}\begin{split}
    \Xi_3(r_+)&=\frac{64\pi G Q^2l^{2n-4}(r_+^2+l^2)^{1-n}r_+^{2n-d-6}}{(d-3)(d-1)(d-2)(n-1)\sqrt{\beta}(r_+^{2d-4}+4\beta Q^2)}\bigg[\frac{2Q(d-2)\sqrt{\beta}}{r_+^{-d-2}(r_+^2+l^2)^{-1}}\\&+r_+^4((d-3)r_+^2+(d-2n+1)l^2)(r_+^{2d-4}+4\beta Q^2)\arctan{\bigg(\frac{2Q\sqrt{\beta}}{r_+^{d-2}}\bigg)}\bigg], \label{32} \end{split}
    \end{equation}
    and
    \begin{equation}\begin{split}
    \Xi_4(r_+)&=\frac{128\pi G Q^2l^{2n-4}r_+^{2n-4}(d^2-2d-3-Qd+Q)}{(d-3)(d-1)^2(n-1)(l^2+r_+^2)^{n-1}(r_+^{2d-4}+4\beta Q^2)^2}\bigg[\big((2d-5)r_+^2\\&+l^2(2d-2n-1)\big)r_+^{2d-4}-4\beta Q^2\big(r_+^2+(2n-3)l^2\big)\bigg]. \label{33} \end{split}
    \end{equation}
    Hence, by using Eq. (\ref{29})-(\ref{33}) along with the expressions of entropy and temperature in (\ref{28}), we can get heat capacity as
    \begin{eqnarray}\begin{split}
    C_Q&=\frac{(n-1)\Sigma_{d-2}(r_+^2+l^2)r_+^{d-2n+1}\chi(r_+)}{4Gl^{2n}(d-2n+2)(d-2n+4)\sqrt{\beta}(n-1)(d-1)(d-2)}\\&\times\frac{\sqrt{\beta}(n-1)(d-1)(d-2)(r_+^2+l^2)^{n-2}(\Delta_1+\Delta_3)+16\pi Gr_+^{2n-3}l^{2n-4}\Delta_2}{\bigg(2r_+^{-2}+\Xi_1(r_+)+\Xi_2(r_+)+\Xi_3(r_+)+\Xi_4(r_+)\bigg)},\label{34}\end{split}
    \end{eqnarray} 
    where
     \begin{eqnarray}\begin{split}
   \Delta_1=\frac{(d-1)r_+^{2n-3}l^{2n-4}}{(n-1)(r_+^2+l^2)^{n-2}}\bigg[\bigg(\frac{1}{r_+^2}+\frac{1}{l^2}\bigg)^{n-1}+\frac{16\pi G H(d-2)}{(d-1)r_+^d}\bigg]-\frac{2}{r_+},\label{35}\end{split}
    \end{eqnarray} 
     \begin{eqnarray}\begin{split}
    \Delta_2=\frac{1}{\sqrt{\beta}}\log{\bigg(1+\frac{4\beta Q^2}{r_+^{2d-4}}\bigg)}-\frac{4Q^2}{(d-3)r_+^{d-2}}\arctan{\bigg(\frac{2Q\sqrt{\beta}}{r_+^{d-2}}\bigg)},\label{36}\end{split}
    \end{eqnarray}
     \begin{eqnarray}\begin{split}
     \Delta_3=\frac{128\pi G Q^2 l^{2n-4}r_+^{2n-3}(Q(d-1)-(d-3)(d+1)}{(d-3)(d-1)^2(n-1)(r_+^2+l^2)^{n-2}(r_+^{2d-4}+4\beta Q^2)},\label{37}\end{split}
     \end{eqnarray}
     and 
      \begin{eqnarray}\begin{split}
     \chi_1(r_+)&=l^2(d-2n+4)\big(r_+^2d+l^2(d-2n+2)\big)F_1\bigg[1,\frac{d}{2},\frac{d-2n+4}{2},-\frac{r_+^2}{l^2}\bigg]\\&-2d r_+^2(r_+^2+l^2)F_1\bigg[2,1+\frac{d}{2},3-n+\frac{d}{2},-\frac{r_+^2}{l^2}\bigg].\label{38}\end{split}
     \end{eqnarray}
     
     \begin{figure}[h]
     	\centering
     	\includegraphics[width=0.8\textwidth]{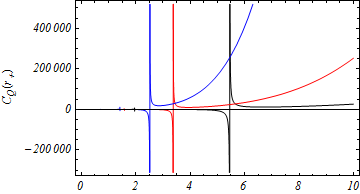}
     	\caption{Heat capacity versus $r_+$. The parameters are selected as $Q=10$, $\beta=0.1$, $\Sigma_{d-2}=100$, $l=1$, $H=1$, $n=2$, $d=5$ (black), $n=2$, $d=6$ (red) and $n=3$, $d=7$ (blue).}\label{skr3}
     \end{figure}
  \begin{figure}[h]
 	\centering
 	\includegraphics[width=0.8\textwidth]{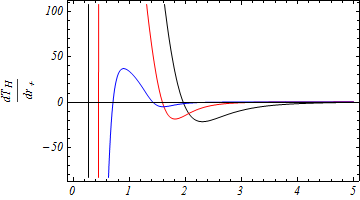}
 	\caption{$\frac{dT_H}{dr_+}$ versus $r_+$. The parameters are selected as $Q=10$, $\beta=0.1$, $\Sigma_{d-2}=100$, $l=1$, $H=1$, $n=2$, $d=5$ (black), $n=2$, $d=6$ (red) and $n=3$, $d=7$ (blue).}\label{skr4}
 \end{figure}
 
    The heat capacity is significant as it plays a crucial role in local thermodynamic stability of black holes. Fig. \ref{skr3} shows the plot of heat capacity $C_Q$ dependent on $r_+$ for suitably chosen values of the parameters $Q$, $\beta$, $H$ and $d$. Note that, the region where it is positive (negative) indicates the local thermal stability (instability) of the black hole. The values of $r_+$ for which it changes sign represents the phase transitions of type I. However, those values of $r_+$ at which it is infinite indicate the existence of phase transitions of type II. Note that, these transitions can also be analyzed from  Fig. \ref{skr4} because the points at which $dT_H/dr_+$ vanishes implies the heat capacity is divergent at these points. Hence, those values at which the curve of $dT_H/dr_+$ crossed the horizontal axis correspond to the phase transitions of type II.

\section{Magnetized hairy black holes of general Lovelock-Scalar gravity} 
The Lovelock polynomial which provides the metric function for magnetized hairy black holes in general LTG can be obtained from (\ref{14a}) as 
\begin{eqnarray}\begin{split}
&\sum_{p=0}^{n-1}\frac{(d-1)!a_p}{(d-2p-1)!}\bigg(\frac{\alpha-f(r)}{r^2}\bigg)^p=\frac{16\pi G M(d-1)}{(d-2)\Sigma^{(\alpha)}_{d-2}r^{d-1}}+\frac{16\pi G(d-1)H}{r^d}\\&+\frac{64Q \pi Gr^{2-d}}{\sqrt{\beta}(d-2)(d-3)}\arctan{\bigg(\frac{2Q\sqrt{\beta}}{r^{d-2}}\bigg)}-\frac{16\pi G}{\beta (d-2)(d-1)}\log{\bigg(1+\frac{4\beta Q^2}{r^{2d-4}}\bigg)}\\&-\frac{128Q^2d\pi G}{(d-3)(d-1)r^{2d-4}} F_1\bigg[1,\frac{d-3}{2d-4},\frac{3d-7}{2d-4},-\frac{4\beta Q^2}{r^{2d-4}}\bigg]. \label{39}\end{split}
\end{eqnarray}
Note that, we have used the scalar configuration in the form (\ref{12}) along with the assumption of pure magnetic field. The expression for finite mass of the hairy black hole associated to the above polynomial takes the form
\begin{eqnarray}\begin{split}
M&=\frac{\Sigma^{(\alpha)}_{d-2}}{16\pi G (d-1)}\bigg[\sum_{p=0}^{n-1}\frac{(d-1)!a_p\alpha^p}{(d-2p-1)!r_+^{-(d-2p-1)}}-\frac{16\pi G(d-1)(d-2)H}{r_+}\\&+\frac{16\pi Gr_+^{d-1}}{\beta(d-2)(d-1)}\times\log{\bigg(1+\frac{4\beta Q^2}{r_+^{2d-4}}\bigg)}-\frac{64\pi GQ^2r_+\arctan{\bigg(\frac{2Q\sqrt{\beta}}{r_+^{d-2}}\bigg)}}{\sqrt{\beta}(d-2)(d-3)}\\&+\frac{128\pi GQ^2d}{(d-3)(d-1)r_+^{d-3}}F_1\bigg(1,\frac{d-3}{2d-4},\frac{3d-7}{2d-4},-\frac{4\beta Q^2}{r_+^{2d-4}}\bigg)\bigg]. \label{40}\end{split}
\end{eqnarray}

Similarly, the Hawking temperature can be computed as 
\begin{eqnarray}\begin{split}
T_{H}(r_+)&=\frac{1}{4\pi Z(r_+)}\bigg[\sum_{p=0}^{n-1}\frac{(d-1)!a_p\alpha^p}{(d-2p-2)!r_+^{2p+1}}+\frac{16\pi G H(d-1)(d-2)}{r_+^{d+1}}\\&+\frac{16\pi G}{(d-2)r_+\beta}\log{\bigg(1+\frac{4\beta Q^2}{r_+^{2d-4}}\bigg)}-\frac{64\pi G Q^2\arctan{\bigg(\frac{2Q\sqrt{\beta}}{r_+^{d-2}}\bigg)}}{\sqrt{\beta}(d-2)(d-3)r_+^{d-1}}\\&+\frac{128\pi G Q^2\big(Q(d-1)-(d+1)(d-3)\big)}{(d-1)(d-3)r_+(r_+^{2d-4}+4\beta Q^2)}\bigg], \label{41}\end{split} 
\end{eqnarray}
where $Z(r_h)$ is defined by
\begin{equation}
Z(r_h)=\sum_{p=0}^{n-1}\frac{pa_p\alpha^{p-1}(d-1)!}{(d-2p-1)!r_+^{2p}}.\label{42}
\end{equation}
Again by employing Wald's method, it is straightforward to obtain entropy as
\begin{equation}
S=\frac{(d-2)\Sigma^{(\alpha)}_{d-2}}{4G}\sum_{p=1}^{n-1}p\alpha^{p-1}\bigg(\frac{(d-2p-1)!a_pr_+^{d-2p}}{(d-2)!}+\frac{b_p(d-3)!N^{d-2p}}{(d-2p)!}\bigg). \label{43}.
\end{equation}

Hence, by using the Eqs. (\ref{41})-(\ref{43}), one can obtain the heat capacity in the form as follows:
\begin{eqnarray}\begin{split}
C_Q&=\frac{\Sigma_{d-2}^{(\alpha)}\sum_{p=1}^{n-1}p\alpha^{p-1}(d-2p)!a_pr_+^{d-2p-1}}{4G(d-1)!\bigg(A(r_+)+\frac{16\pi G H(d-1)(d-2)}{r_+^{d+1}}+\sum_{p=0}^{n-1}\frac{(d-1)!a_p\alpha^p}{(d-2p-2)!r_+^{2p+1}}\bigg)^{-1}}\\&\times\bigg[\bigg(\frac{dA}{dr_+}-\sum_{p=0}^{n-1}\frac{(d-1)!a_p\alpha^p(2p+1)}{(d-2p-2)!r_+^{2p+2}}-\frac{16\pi G H(d^2-1)(d-2)}{r_+^{d+2}}\bigg)-\\&\frac{dZ/dr_+}{Z(r_+)}\bigg(A(r_+)+\frac{16\pi G H(d-1)(d-2)}{r_+^{d+1}}+\sum_{p=0}^{n-1}\frac{(d-1)!a_p\alpha^p}{(d-2p-2)!r_+^{2p+1}}\bigg)\bigg]^{-1}, \label{44}\end{split}
\end{eqnarray} 
\begin{eqnarray}\begin{split}
A(r_+)&=\frac{16\pi G}{(d-2)r_+\beta}\log{\bigg(1+\frac{4\beta Q^2}{r_+^{2d-4}}\bigg)}+\frac{128\pi G Q^2\big(Q(d-1)-(d+1)(d-3)\big)}{(d-1)(d-3)r_+(r_+^{2d-4}+4\beta Q^2)}\\&-\frac{64\pi G Q^2}{\sqrt{\beta}(d-2)(d-3)r_+^{d-1}}\arctan{\bigg(\frac{2Q\sqrt{\beta}}{r_+^{d-2}}\bigg)}, \label{45}\end{split}
\end{eqnarray}   
and
\begin{eqnarray}\begin{split}
\frac{dA}{dr_+}&=\frac{256\pi G Q^2}{(d-3)(d-1)r_+^2(r_+^{2d-4}+4\beta Q^2)^2}\bigg(\big(d^3-(Q+4)d^2+(3Q+2)d\\&+(3-2Q)\big) r_+^{2d-4}+4(d-3)\beta Q^2\bigg)+\frac{64\pi G (d-1)Q^2}{\sqrt{\beta}(d-2)(d-3)r_+^{d}}\arctan{\bigg(\frac{2Q\sqrt{\beta}}{r_+^{d-2}}\bigg)}\\&-\frac{16\pi G}{\beta r_+^2(d-2)}\log{\bigg(1+\frac{4\beta Q^2}{r_+^{2d-4}}\bigg)}. \label{46}\end{split}
\end{eqnarray} 
\section{Summary and conclusion} 

 Lovelock gravity in the presence of nonlinear electrodynamics is a very interesting area of research, which inspired us to study black holes in this context. Therefore, in this paper, I started my work with the dimensionally continued gravity in the presence of double-logarithmic electrodynamics and conformal scalar field. Then, I construct the associated equations of motion and derive the new solution (\ref{15}) for the magnetized hairy black holes. I mainly focused on the magnetized black holes because the pure electric or dyonic cases in double-logarithmic electrodynamics cannot admit the metric function in a closed form. The thermodynamic properties of these hairy black holes are discussed and the exact expressions for finite mass, Hawking temperature and heat capacity are calculated. The satisfaction of generalized first law of thermodynamics has also been verified. The thermodynamic quantities are plotted in different dimensions and their behaviours are discussed. The regions corresponding to local thermodynamic stability can be identified from these plots. The plots of heat capacity and Hawking temperature shows that thermal phase transitions of black holes are also possible. The phase transitions of type I corresponds to those values of $r_+$ at which the specific heat changes sign. On the other side, the type II phase transitions of black holes corresponds to the singularities of heat capacity. Finally, the hairy black holes in Lovelock gravity are briefly investigated within the chosen model of double-Logarithmic electrodynamics.

It is worth noting that when $\beta\rightarrow 0$ both the solution (\ref{15}) and the polynomial (\ref{39}) will describe the Maxwellian charged hairy black holes. The black holes without scalar hair can also be recovered by simply putting either $H=0$ or $b_p's=0$ in (\ref{15}) and (\ref{39}). However, by choosing $Q$ equal to zero, the metric function for neutral black holes can be obtained.

It would be very interesting to study Hawking radiations, thermal fluctuations and grey body factors for the hairy black holes obtained in this paper. In addition to this, one can also use the model of double-logarithmic electrodynamics for the analysis of accelerated expansion of the universe. Similarly, the new black hole solutions of quartic and quintic quasi-topological gravities can also be obtained in the presence of double-logarithmic electrodynamics.

\end{document}